\documentclass[a4paper]{spie}

\usepackage[]{graphicx}
\usepackage{aas_macros}
\usepackage[]{subfigure}
\title{XPOL: a photoelectric polarimeter onboard XEUS}

\author{Enrico~Costa\supit{a}, Ronaldo~Bellazzini\supit{c},
Jean~Bregeon\supit{c}, Alessandro~Brez\supit{c},
Massimo~Frutti\supit{a}, Sergio~Di Cosimo\supit{a},
Luca~Latronicio\supit{c}, Francesco~Lazzarotto\supit{a},
Giorgio~Matt\supit{d}, Massimo~Minuti\supit{c},
Ennio~Morelli\supit{e}, Fabio~Muleri\supit{a,b},
Michele~Pinchera\supit{c}, Massimiliano~Razzano\supit{c},
Alda~Rubini\supit{a}, Paolo~Soffitta\supit{a},
Gloria~Spandre\supit{c} \skiplinehalf \supit{a} Istituto di
Astrofisica Spaziale e Fisica Cosmica, Via del Fosso del Cavaliere
100, I-00133 Roma, Italy;
\\
\supit{b} Universit\`{a} di Roma Tor Vergata, Dipartimento di
Fisica, via della Ricerca Scientifica 1, 00133 Roma, Italy
\\
\supit{c} Istituto Nazionale di Fisica Nucleare, Largo B. Pontecorvo 3, I-56127 Pisa,  Italy
\\
\supit{d} Universit\`{a} di Roma Tre, Dipartimento di Fisica
E.Amaldi, via della Vasca Navale 84, 00146 Roma, Italy
\\
\supit{e} Istituto di Astrofisica Spaziale e Fisica Cosmica, Via
Gobetti 101, I-40129 Bologna, Italy }

\authorinfo{Further author information: (Send correspondence to Enrico Costa)
\\Enrico Costa: E-mail: enrico.costa@iasf-roma.inaf.it,
Telephone: +39-0649934004}

\begin{document}
\maketitle

\begin{abstract}
The XEUS mission incorporates two satellites: the Mirror
Spacecraft with 5 m$^{2}$ of collecting area at 1 keV and 2
m$^{2}$ at 7 keV, and an imaging resolution of 5" HEW and the
Payload Spacecraft which carries the focal plane instrumentation.
XEUS was submitted to ESA Cosmic Vision and was selected for an
advanced study as a large mission. The baseline design includes
XPOL, a polarimeter based on the photoelectric effect, that takes
advantage of the large effective area which permits the study of
the faint sources and of the long focal length, resulting in a
very good spatial resolution, which allows the study of spatial
features in extended sources. We show how, with XEUS, Polarimetry
becomes an efficient tool at disposition of the Astronomical
community.

\end{abstract}


\section{Introduction}

XEUS is an ambitious mission planned to be flown $\sim$55 years
after the start of X-ray astronomy. XEUS focal plane
instrumentation is extremely evolved, especially in the domain of
imaging non-dispersive spectroscopy and of wide field imaging with
a good spectral response. This follows an almost continuous
development from the first rockets through very successful
missions such as Einstein, ROSAT, ASCA, SAX, Chandra, XMM. The
development of polarimeters has not proceeded in parallel. In fact
after the first attempts and the first success with OSO-8, no
polarimeter has been embarked aboard a mission, with the exception
of SPECTRUM-X-Gamma, that never arrived to the launch. Polarimetry
is therefore an all to dig field, and a relatively extended
literature (at least compared with the shortage of data) suggests
that the crop would be highly rewarding. Nowadays new polarimeters
based on the photoelectric effects are available. The INFN of Pisa
has developed the Gas Pixel Detector, in the frame of a
collaboration with IASF\cite{Costa2001, Bellazzini2006,
Bellazzini2007}. These devices combine the capability to measure
the polarization with good imaging and can be employed as focal
plane detectors, allowing for the same dramatic improvement
occurred for imaging with the arrival of Einstein mission. We
remind that Einstein was a step forward also from the point of
view of satellite attitude. In the pioneering satellites,
stabilized on one axis, the detectors had a slat collimator
misaligned with respect to rotation. A source was detected as an
excess of counts following the collimator profile. Einstein and
all the following satellites were stabilized on three axis and a
source was a cluster of events in the image consistent with the
telescope psf. The diffraction polarimeter is non-dispersive and
requires rotation (both of analyzer and of detector) to perform
the measurement. The scattering polarimeter is intrinsically
non-dispersive but requires the rotation of the whole to
compensate huge systematic effects. Since the rotation was no more
provided by the satellite, the polarimeter introduced a serious
complication of the focal plane, to be added to the complication
of swapping from one instrument to the other in the focal plane.
Since the scientific interest of polarimetry was out of question,
these mismatching were the cause of the removal of polarimeters
from the major X-ray missions (Einstein, Chandra), where it was
foreseen in the beginning.

Beside the advantage of being small and working at room
temperature, there is the additional advantage of not requiring
rotation: this removes a further area of mismatching with other
instruments. The change of instruments in the focal plane is
intrinsically and safely resolved by the formation flight
technology itself.

The first obvious answer to the question "Why to include a
polarimeter in the focal plane of XEUS?" could be that it is
simple and does not require large resources.

In fact we could object that XEUS will, in any case, devote a
minor fraction of its time to polarimetry. It is likely that
before XEUS a dedicated mission will be flown. Such a mission
could perform very long pointing of some target of particular
interest and partially compensate with observing time the smaller
effective area.

We will demonstrate in the following that a GPD polarimeter aboard
XEUS can achieve scientific results of high value, that there
results will solve some hot topics which are within the scientific
targets of XEUS and that can be achieved with XEUS only, and not
with a pathfinder mission of lower performance.

%
%

\section{XPOL}

The purpose of the XPOL is to provide, in the energy range 2 -
10keV, polarization measurements simultaneously with angular
measurement (5 arcsec), spectral measurements (E/$\Delta$ E of
$\sim $5 @ 6 keV) and timing at few $\mu$s level.  The FOV is of
1.5 $\times$ 1.5 square arc minutes. XPOL is based on a Gas Pixel
Detector, composed of a gas cell and a VLSI that acts as the
bottom of the detector,  and as Front End Electronics. The signals
from the ASIC chip are delivered to an Interface Electronics,
where are A/D converted and tagged with time. The whole is
controlled by an electronics with a processor that analyzes the
track and creates the telemetry packets. XPOL is provided with a
baffle to prevent the photons from the bright X-ray sky to impinge
on the detector window. A filter wheel carries calibration sources
to check the stability of the detector performance with time and
filters to allow for particular strategies to observe very bright
sources.

\subsection{The Detector} \label{sec:Detector}
The heart of the polarimeter is the Gas Pixel Detector. It is a
counter, with a beryllium window 50 $\mu$m thick, filled with a
mixture of low atomic number components (usually He 20$\%$ DME
80$\%$). The photon is converted in an absorption/drift gap 10 mm
thick. The photoelectron interacts with atoms close to the impact
point ad is slowed by ionization and scattered by the field of
nuclei. The result is a track of electron-ion pairs. The electrons
in the track are drifted by a constant electric field to a Gas
Electron Multiplier, a polyimide film, metal coated on both sides,
with a matrix of holes on an hexagonal pattern, with a pitch of 50
$\mu$m. Each hole multiplies in a proportional way the charge.
Therefore the track is amplified, while preserving the information
on the shape and on the charge. Multiplied electrons are collected
by a plane of metal pads, close to the GEM, also with hexagonal
pattern and with the same pitch of 50 $\mu$m. Each pad is the
input of a complete electronic chain that detects the charge. Pads
and front end electronics are part of a VLSI chip, based on 0.18
$\mu$m CMOS technology. The chip has the capability to self
trigger and fetches at the output only the content of a Region of
Interest, including the pixels that triggered. Since the chip has
a total of 105600 pixels, and a track typically produces a charge
on 50-100 pixels (depending on the energy of the photon), this
design prevents the divergence of dead-time that would be needed
to read the whole detector image. The analysis of the tracks
allows to derive the impact point (with a precision one order of
magnitude better than that of the centroid of the charge) and the
ejection direction of the primary photoelectron. The latter
carries the information of the polarization of the beam. The
precision on the impact point is of the order of  $\sim$ 150
$\mu$m FWHM, largely oversampling the PSF but this is not
completely exploited because of the blurring due to the absorption
of photons from an inclined beam at different heights in the gas.
This last effect is determining the actual resolution of XPOL.

The GPD detector ad its polarimetric capabilities have been
extensively described elsewhere. Below we give for XPOL figures of
sensitivity which are based on experimental data on existing
prototypes, without including any margin for the possible (and
foreseen) improvements\cite{Bellazzini2006, Muleri2008}.

\begin{figure}[htbp]
\begin{center}
\includegraphics[angle=0,width=16cm]{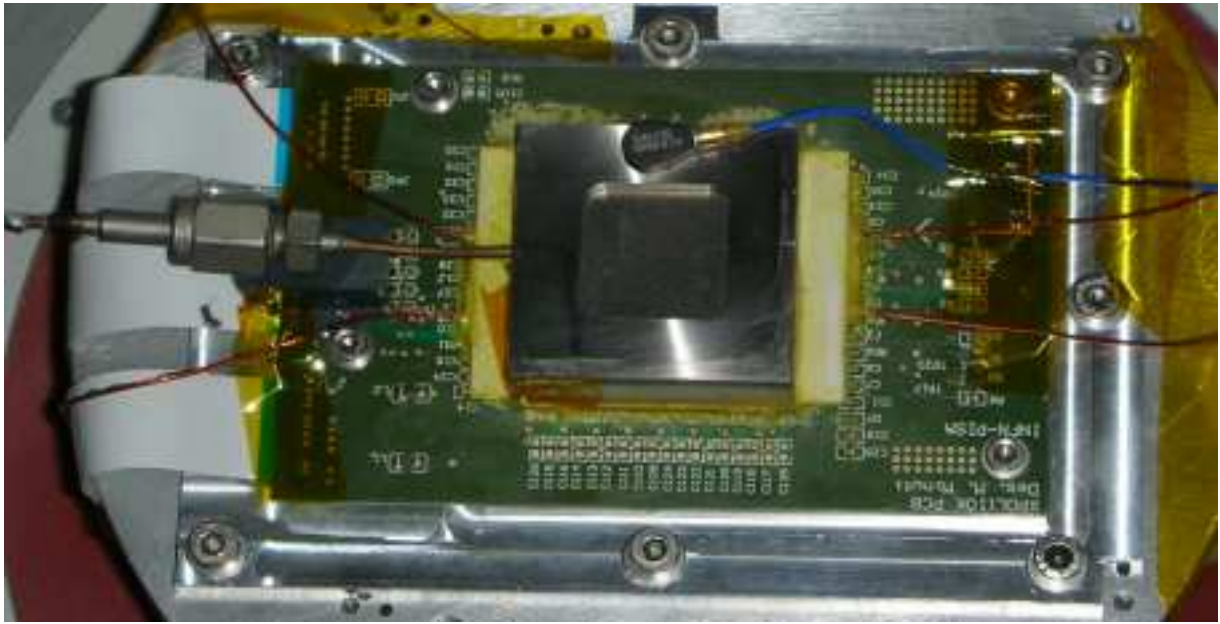}
\end{center}
\caption{A prototype Gas Pixel detector in the facility for
vibration test\label{fig:DetVib}}
\end{figure}

The level of readiness of the detector is in a good shape. Sealed
prototypes, built with low  desorption materials, have been tested
for more than one year without any evidence of change. It should
be considered that the technology for the manufacture of long
duration gas cells for proportional counters and GSPCs to be
employed in the space, is very well established. The stability of
the mixture has been tested. Further testing for the robustness of
the GEM to spark in presence of ions will be performed in a short
time. Anyway it should be considered that the GEM is operated at a
very low gain level ($\sim$500) that results in a condition much
safer than that of other gas multiplication devices. A point to be
clarified is the capability to handle the huge flux of data
deriving from the observation of bright sources with XPOL (up to
20000 counts/second). With relatively minor changes to the ASIC
chip, that will not impact on the noise figure,  we think to
arrive to this result. In Fig.~\ref{fig:DetVib} we show a
prototype detector subject to vibration testing.

\subsection{The Focal Plane} \label{sec:focalplane}
\begin{figure}[htbp]
\begin{center}
\includegraphics[angle=0,width=8cm]{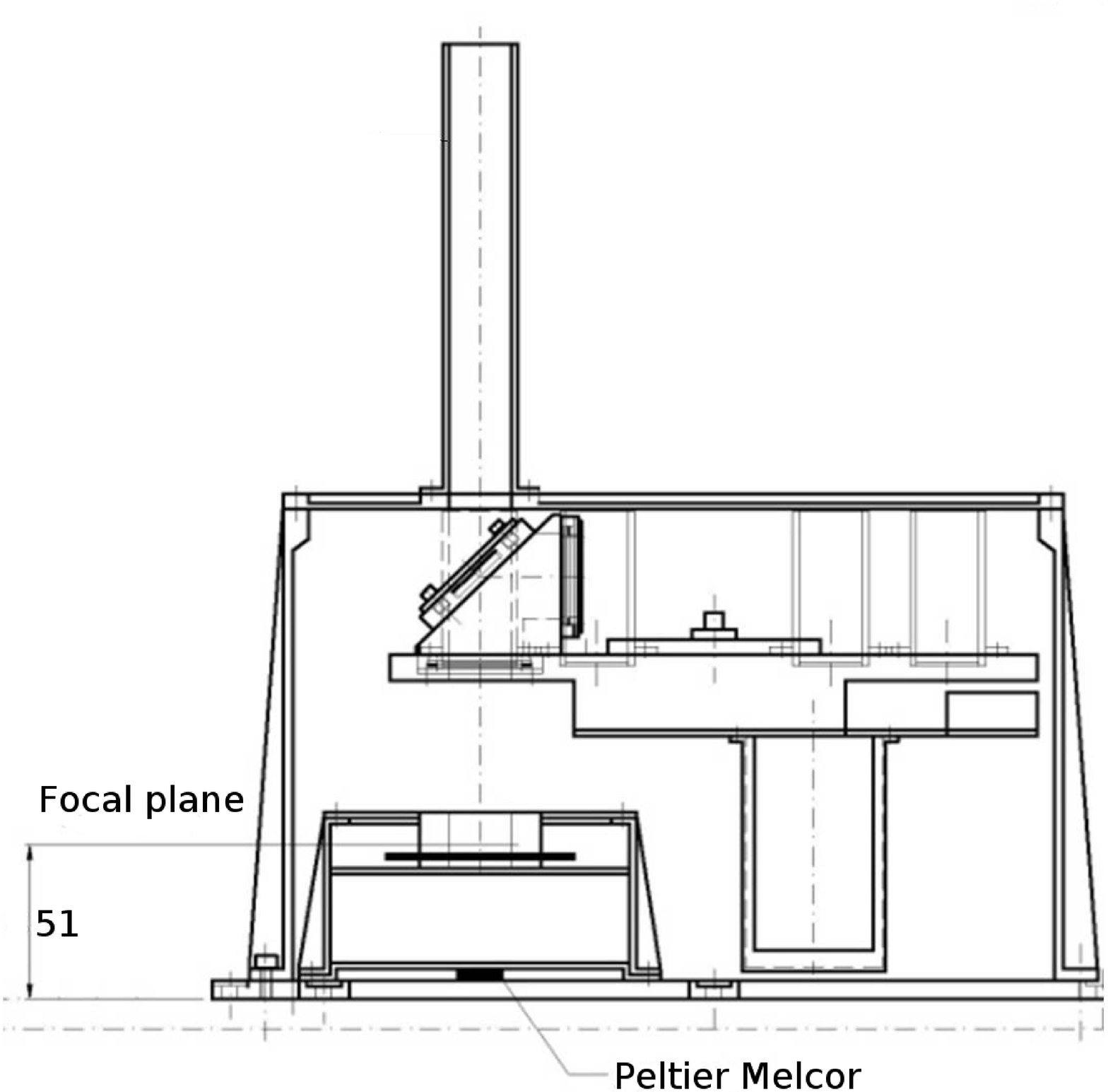}
\end{center}
\caption{The focal plane of XPOL including the detector, the
back-end electronics, the filter wheel and the baffle (the
dimensions of the baffle are not representative)
\label{fig:FocalPlane}}
\end{figure}
According to different operative modes a filter wheel will control
and determine radiation impinging on the detector. The filter
wheel is a disc with different positions on it that can be rotated
into the optical field of view with a motor. The Filter wheel
position may be selected for observation, for calibration source
deployment or for safety. Safety operations may be required
autonomously or via ground control to prevent excessive charged
particle flux (solar flare or local magnetospheric storms).

The filter wheel is foreseen to have 6 positions:
\begin{itemize}
\item Position A) Closed  (Operative mode: power-off, stand-by,
electrical calibration, observation*) \item Position B)  Opened
(Operative mode : Observation, normal rate ). \item Position C)
With transmission filter (Operative mode: observation, very high
rate). \item Position D) With diaphragm (Operative mode :
Observation, small FOV). \item Position E) With calibration source
I (Fe55). Operative mode : calibration. \item Position F) With
calibration source Ti/Cd-109 (TBC) . Operative mode : calibration.
\end{itemize}

The presence of calibration sources aboard will be useful to
monitor the stability of the gain and of the response to polarized
photons. Therefore we foresee an source providing unpolarized
photons and another one emitting polarized ones (by bragg
diffraction). They will be put periodically in the front of the
detector window.

A baffle of carbon fiber with thin metal plating will prevent the
direct vision of the diffuse X-ray background from the sky. The
dimension of the baffle are still to be defined, on the basis of
the extension of the skirt that will surround the optics
satellite.

The Detector, the back-end electronics and the filter wheel will
be enclosed within a protective carter which will also support the
baffle. In order to stabilize the gain the detector will have its
own thermal control. In Fig.~\ref{fig:FocalPlane} we show the
focal plane. The dimensions of the baffle are not representative.

\subsection{The Electronics} \label{sec:electronics}

The back-end electronics is a small box connected with the
detector with flexi cables at a distance of around 20 cm. It
includes the logics to program the ASIC chip, the logic to read
the signal from the chip, to A/D convert them, to flag with time
and to transfer to the control electronics. All the logic
functions are performed by an FPGA. Near to the detector there
will be also High Voltage Power Supplies.

The control electronics is a box that can be placed also at a
certain distance from the focal plane consists of:
\begin{itemize}
\item DC/DC converters to provide stabilized low voltages \item A
DSP processor \item The Mass Memory \item the housekeeping
conditioner
\end{itemize}
The  DPU of the control electronics programs the back-end
electronics, receives the packets of events and organizes for
telemetry or, alternatively, for storage in the mass memory. Since
the amount of data produced by the polarimeter is high we are
selecting the processor with the requirement that it is capable to
perform on-board the analysis of the tracks.

Operating modes foreseen are:
\begin{itemize}
\item Electric calibration mode of pedestals (filter A = door
closed)
\item Electric calibration mode by test pulse (filter A =
door closed)
\item Calibration with radioactive source Fe55
(filter E)
\item Calibration with radioactive source mixed (filter
F)
\end{itemize}
The science modes are:
\begin{itemize}
\item Normal (Filter B = all open ). No post-processing Diaphragm
\item Diaphragm (Filter D = f.o.v partially covered). No
post-processing \item High rate (Filter B = all open ).
Post-processing
\item Extremely high rate (Filter C = all field
attenuated) Post-processing
\end{itemize}

All these science operative modes are the same from the point of
view of detectors and FEE configuration, time tagging, A/D
conversion and zero suppression. They differ for the strategy to
avoid overwhelming the mass memory in case bright sources are
observed. In the normal mode data after zero suppression, the
track image, are stored to the Mass Memory to be further forwarded
to telemetry. In high rate, when the XPOL observation is over and
the telescope is allocated to another instrument, data are
recovered from the Mass Memory, compressed by DPU with onboard
analysis of polarization (position, time and angle) and stored
again in Mass Memory. In case the target source is faint and
another stronger source is present in the field of view, the
latter can be removed by the use of a diaphragm: this is the
diaphragm mode. In case of an extremely bright source, that could
exceed the capability of data handling, all the field will be
attenuated with a filter C.

The XPOL MM is dimensioned (16 GByte) to store data from a 5000 s
observation of a 1 Crab source. The same function could be
performed on the P/L Mass Memory provided that it is made
available for the time needed for post-processing. The processing
time will normally be <20 times the acquisition time. The data
flow for a very bright source could arrive to $\sim$ 28 Mbit/s for
a total  memory occupation of 16GB.The processing to compress data
could take $\sim$ 5 days. After compression it would reduce to
$\sim$ 1 GByte, which could be downloaded over subsequent
telemetry windows at suitably lower rate interleaved with normal
science data, or special communications windows requested for this
download. This means that after the pointing of a very bright
source XPOL cannot be operated for around 2 days.
\begin{figure}[htbp]
\begin{center}
\includegraphics[angle=0,width=17cm]{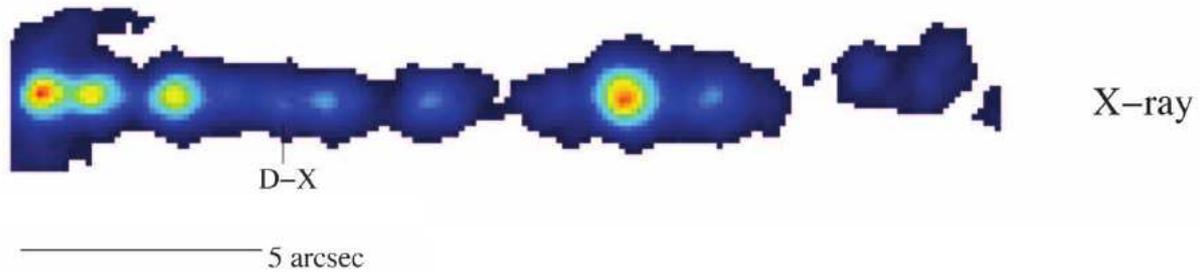}
\end{center}
\caption{The X-ray image of the M87 jet by Chandra. XPOL can
measure the polarization of the brightest knot down to
5$\%$\label{fig:M87}}
\end{figure}
\begin{figure}[htbp]
\begin{center}
\includegraphics[angle=0,totalheight=18cm]{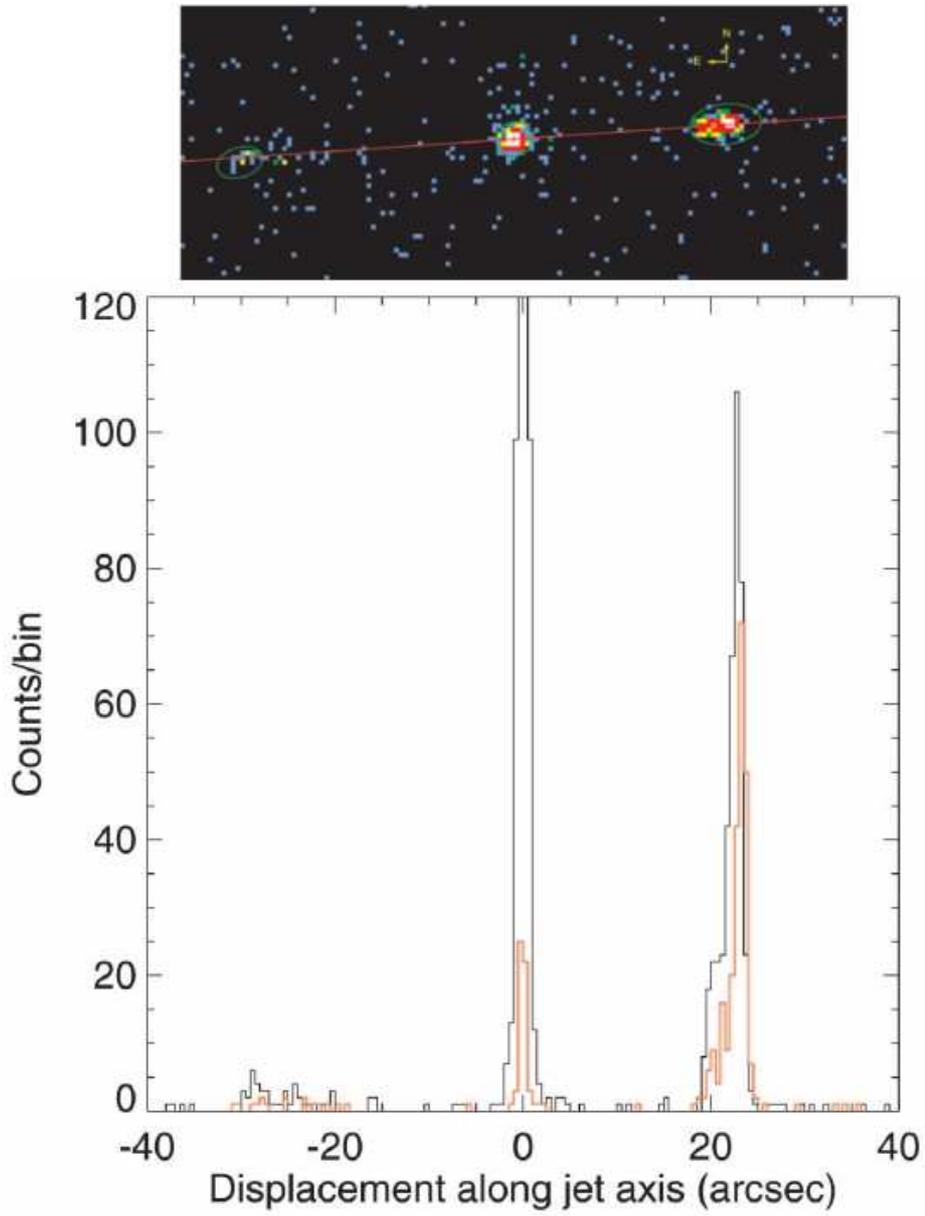}
\end{center}
\caption{The very faint jet of the micro-quasar J1550-564. The
Minimum Detectable Polarization for one day observation with XPOL
is 14$\%$ \label{fig:J1550}}
\end{figure}
\begin{figure}[htbp]
\begin{center}
\includegraphics[angle=90,width=16cm]{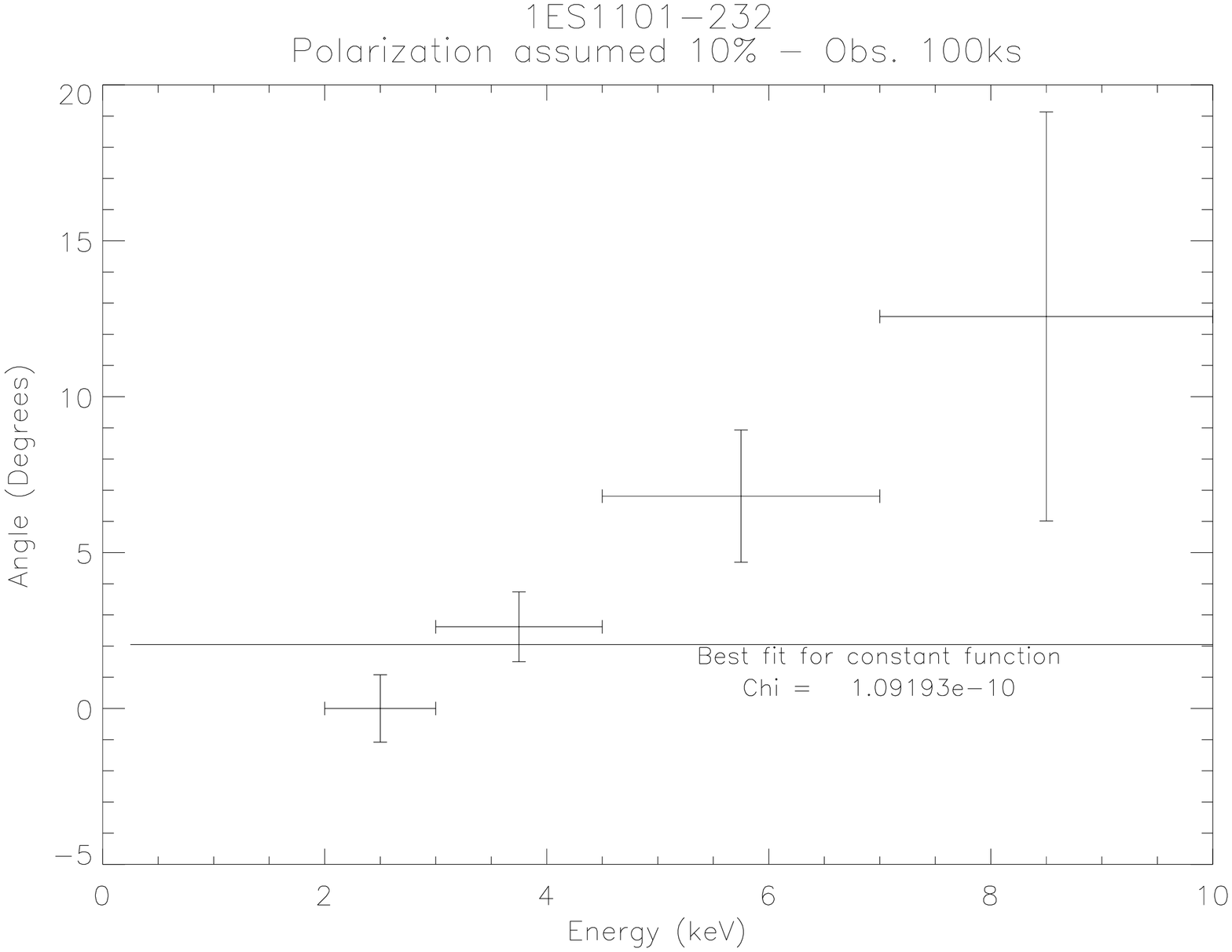}
\end{center}
\caption{The minimum variation of the polarization angle of blazar
1ES1101-232 (z=0.186) detectable by XPOL in one day at 3$\sigma$
level. \label{fig:QGplot}}
\end{figure}

\section{Scientific Performances} \label{sec:performances} The science
rationale of XEUS is built on three major topics:
\begin{itemize}
\item Co-evolution of galaxies and their supermassive black holes
\item Evolution of large scale structure and nucleosynthesis \item
Matter under extreme conditions
\end{itemize}
Moreover XEUS will be open to the community to face many other
scientific items as an observatory. Polarimetry will mainly
contribute to the study of matter in extreme conditions. The main
contribution of XEUS will be in the study of matter under extreme
conditions.

One of the most interesting target for XEUS is the study of the
effects of strong gravitation fields on the radiation, as
predicted by General Relativity. The matter accreting on a compact
object (Neutron Star or Black Hole) is organized in an accretion
disk. Due to the high asymmetry of such a system the radiation
emitted or scattered by the disk will have a certain degree of
polarization. In 1960 Chandrasekhar\cite{Chandrasekhar60} had
computed that the thomson scattering in an infinitely flat cloud
will produce a polarization parallel to the major axis of the
projection of the �disk�in the sky. It will never exceed the limit
of 11.7 $\%$. Later Sunyaev and Titarchuk\cite{SunyaevTitarchuk85}
demonstrated that if the X-ray emission from an accretion disk is
produced by the Comptonization of low frequency radiation, a very
high degree of polarization can be reached for the hard radiation.
Polarization can be negative (parallel to the disk axis) or
positive (perpendicular to the disk axis). In any case, for
reasons of symmetry the photons will be polarized perpendicular or
parallel to the disk. But in the path to the observer the
radiation will experience the strong gravitational field from the
central object. If this is a Black Hole the effect in the observer
frame will be observed as a rotation of the polarization angle.
Stark  Connors and Piran\cite{Stark1977, Connors1980} computed the
effect for the case of a galactic black hole in a binary system.
Since the photons of higher energy are emitted close to the BH,
the rotation will be more effective at higher energies. This
effect of rotation of polarization angle with energy is a unique
signature of the presence of a Black Hole. Moreover the dependence
of polarization amount and angle on energy will be different for
Kerr and Schwartschild black holes. This is one of the most
powerful probes of gravitation near the BH horizon. The
capabilities of XPOL to perform such a test on Cyg X-1 have been
shown by Bellazzini et al.\cite{Bellazzini2006b}.

Another hot topics of high energy astrophysics is the structure
and physics of jets. These are mainly observed by radio telescopes
that provide both high resolution imaging and polarimetry. But in
order to study the structure of regions of freshly accelerated
electrons and improve the insight on the acceleration mechanisms
themselves the X-ray imaging are a fundamental tool. Likely X-ray
polarimetry will single out the formation of plasmoids by time
resolved polarimetry of the central object, but XEUS, with its
large collecting area and with its excellent angular resolution
will give the opportunity to perform angular resolved polarimetry
of knots of brighter jets. In Fig.~\ref{fig:M87} we show the X-ray
structure of M87 as detected by Chandra\cite{PerlmanWilson05}.
With an observation of 10$^{5}$s XPOL can measure the polarization
of knot A down to the level of 5$\%$. It is also apparent that the
angular resolution of 5 arcseconds is essential for such a
measurement. Also the very faint knot of the galactic micro-quasar
XTE J1550-564\cite{kaaret03} can be observed by XPOL with a
Minimum Detectable Polarization of 14$\%$.

Last but not least we want to mention the capability of XPOL to
test theories of Quantum Gravity. The so called Loop Quantum
Gravity predicts that at very long timescale a violation of
Lorentz invariance occurs. The two states of circular polarization
have a different velocity and this difference increases with
energy. Since also the wave-number is proportional to the energy
of the radiation, the total effect is a rotation of the
polarization plane with the distance and with the square of the
energy\cite{Gambini1999}. The amount of this effect of
birefringence is unknown and only upper limits are there. But
since this is one of the few ways to derive experimental
information about QG theories, the continuous search for more
sensitive measurements (even of upper limits) is a worthwhile
task. X-ray is the highest energy band where sensitive polarimetry
of sources at cosmologic distances can be performed. We assume
that Blazars are good candidates to have a high degree of
polarization with angle independent on the energy (at least within
a decade). If this hypothesis is verified on nearby blazars (in
the synchrotron regime) we move to far-away blazars and search for
a rotation of polarization angle with the energy proportional to
the distance. In Fig.~\ref{fig:QGplot} we show that XPOL is
capable to reject at 3$\sigma$ the hypothesis of constant angle
with a 100ks observation of blazar 1ES1101-232 if the coupling
constant is 1$\times 10^{-10}$. This would improve of 5 orders of
magnitude the previous upper limit.

\begin{figure}[htbp]
\begin{center}
\includegraphics[angle=90,width=14cm]{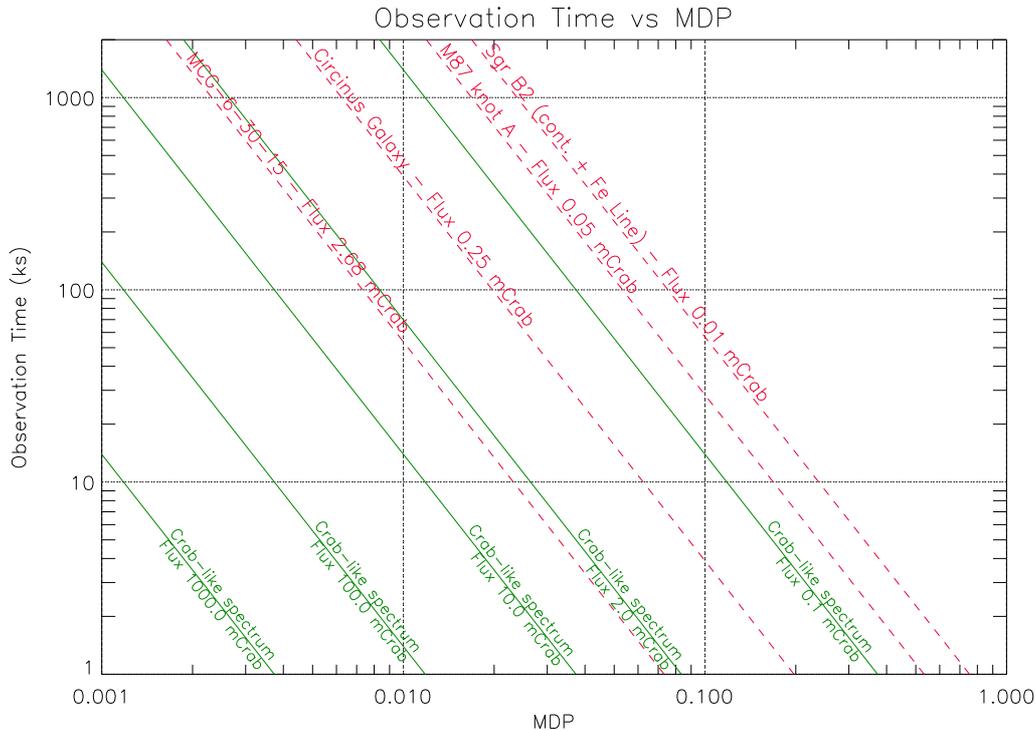}
\end{center}
\caption{The Minimum Detectable Polarization as a function of
observing time for a few representative sources
\label{fig:XEUSsens}}
\end{figure}

\section{Conclusions}

XPOL aboard XEUS is capable to perform some measurements which are
a significant step forward in High Energy Astrophysics and that
cannot be performed by none of the various lower profile proposed
missions. In Fig.~\ref{fig:XEUSsens} we show the time needed to
achieve a certain level of Minimum Detectable Polarization with
XPOL.

XEUS is unique mainly under two respects:

\begin{itemize}
\item The collecting area two orders of magnitude larger than any
dedicated mission \item The angular resolution of few arcseconds
that derives from the long focal length. With such a length also
the blurring due to the finite thickness of the detector is not
very effective.
\end{itemize}
How would it compare with pathfinder missions? XEUS could
reasonably dedicate to polarimetry only a fraction of its time
(let us say 1/10) while a pathfinder could perform full time
polarimetry. The step in surface to have a drastic improvement
with respect to pathfinders is of two orders of magnitude.(namely
the area should be of $\approx5m^{2}$). Both these parameters are
subject to a potential reduction in the frame of design trade-off
in order to decrease costs or weights or, simply, to cope the
performance of the actual optics technology. A decrease of
collecting surface $R_{s}$ results in a proportional increase of
the observing time: t$\longrightarrow$ t/$R_{s}$. Or In a
reduction of MDP as $R_{s}^{1/2}$ for the same observing time e.g.
this would result in:
\begin{itemize}
\item a reduced sample of AGN, with a poorer coverage of parameter
space \item a significant loss of sensitivity to variability of
polarization angle with time (namely on testing strong gravity in
extragalactic BHs).
\end{itemize}

On the other side, since the source will still exceed the
background  a relaxation of the angular resolution would not
impact on polarimetric sensitivity. but Would miss a few topical
targets that only XEUS can do. The most important: polarimetry of
all details of the Crab and of other Pulsar Wind Nebulae,
polarimetry of jets (galactic and extragalactic), polarimetry of
bright knots of shell-like SNR, fast variability of polarization
angle, due to General Relativity effects, in AGN.

\section*{Acknowledgments}
The authors acknowledges financial support from Agenzia Spaziale
Italiana (ASI).

\bibliography{References}   
\bibliographystyle{spiebib}   

\end{document}